\documentclass{article}

\usepackage{arxiv}

\usepackage[utf8]{inputenc} 
\usepackage[T1]{fontenc} 
\usepackage{hyperref}  
\usepackage{url} 
\usepackage{booktabs}  
\usepackage{amsfonts}  
\usepackage{amsmath}  
\usepackage{nicefrac}   
\usepackage{microtype} 
\usepackage{lipsum}
\usepackage{graphicx}
\usepackage[dvipsnames]{xcolor}

\usepackage{etoc}
\usepackage{enumitem}
\usepackage{float}

\usepackage{comment}

\usepackage[
    backend=biber,
    style=numeric-comp,
    sorting=none, 
    eprint=false,
    isbn=false,
    url=false,
    doi=true
]{biblatex}
\usepackage{titlesec}
\titlespacing*{\subsubsection}
{0pt}{.1ex plus .1ex minus .1ex}{.1ex plus .1ex}

\usepackage{etoc}

\title{Information Theory of Random Composition Trees}

\usepackage{authblk}

\author[1,*]{Doron Sivan}
\author[2]{Weishun Zhong}
\author[3]{Tankut Can}
\author[1,2]{Mikhail  Katkov}
\author[1,2,$\dagger$]{Misha Tsodyks}

\affil[1]{Department of Brain Sciences, Weizmann Institute of Science, Rehovot, 76100, Israel}
\affil[2]{School of Natural Sciences, Institute for Advanced Study, Princeton, NJ, 08540, USA}
\affil[3]{Department of Physics, Emory University, Atlanta, GA 30322, USA}

\begin{document}
\maketitle

\footnotetext[1]{doron.sivan@weizmann.ac.il}
\footnotetext[2]{mtsodyks@gmail.com}

\begin{abstract}
Random composition trees model the hierarchical semantic structures of meaningful texts as a statistical ensemble generated by recursively splitting texts into coherent chunks.
A recent study shows that different text types can be described by random composition trees with distinct branching factors, and that the entropy rates of these trees exhibit a similar dependence across text types as the entropy rates of the actual texts. 
Motivated by this correspondence, we analyze information-theoretic properties of this stochastic splitting model. 
First, we demonstrate that the entropy of random composition trees grows linearly with the root size, and we derive an analytical expression for the entropy rate as a function of the branching factor. 
Second, we prove that large trees exhibit typicality: with high probability, they belong to a typical set in which each tree is approximately equally likely.
Our analysis reveals that the random composition tree model shares two key information-theoretic properties of language when modeled as a stationary and ergodic stochastic process: linear entropy growth and typicality.
\end{abstract}

\thispagestyle{plain}

\etocdepthtag.toc{mtoc} 
\etocsettagdepth{mtoc}{section} 
\etocsettagdepth{atoc}{none}    

\section{Introduction: The Random Composition Tree Model}
Meaningful texts can be described by a hierarchical semantic structure. 
Such a structure is also hypothesized to underlie the way people comprehend a text \cite{Kintsch1998}, and the ability to summarize the text at different levels of abstraction \cite{Brown1983}.

\begin{figure}[htbp]
    \centering
    \includegraphics[width=\textwidth]{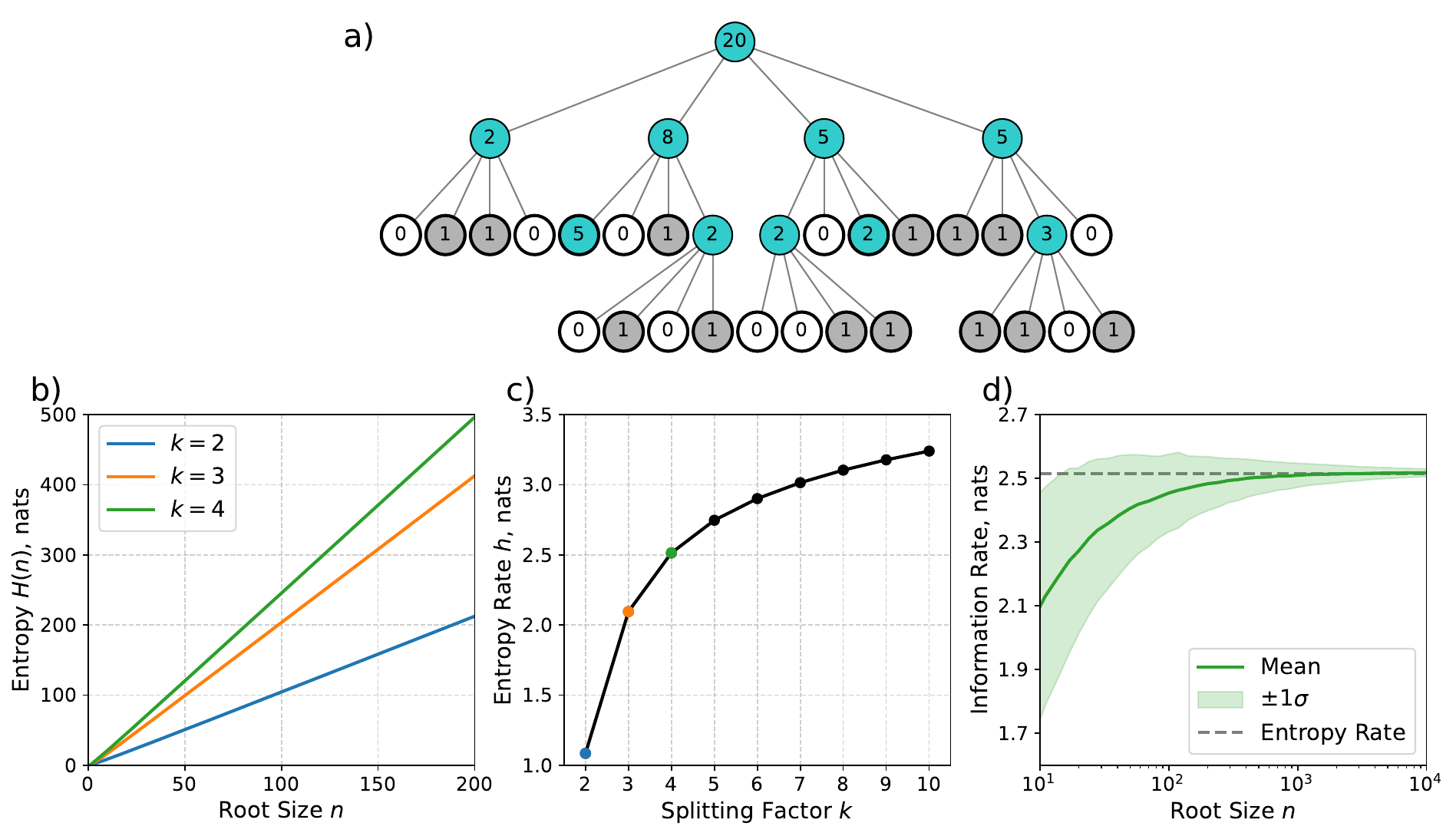}
    \caption{Entropy of random composition trees. a) A realization of a random composition tree with branching factor $k=4$ and root size $n=20$ (nodes of size zero in white, size one in grey, else in aqua; leaf nodes have a thicker boundary). b) The entropy of random composition trees grows linearly with the size of the root, for different values of the branching factor $k$. c) The asymptotic entropy rate of random composition trees as a function of the branching factor $k$. d) The information rate of large trees concentrates sharply around its expected value, which is the entropy rate.}
    \label{fig:tree_numerics}
\end{figure}

In two recent papers \cite{Zhong2025,Zhong2026}, the semantic structure has been modeled as a tree that is created by recursively splitting a text into coherent parts, forming chunks.
Each chunk is thought to be an abstract representation, a concise summary, of its corresponding text segment.
Following \cite{Zhong2025}, we model the semantic structures of texts of a given length as a statistical ensemble of trees generated by a recursive stochastic splitting rule: we split each chunk into $k$ (possibly empty) chunks by placing $k-1$ divisions at random positions between the text units of the chunk.
By labeling each chunk with its size, i.e., the number of text units it summarizes, each splitting constitutes a (weak) composition: an ordered sequence of $k$ non-negative integers that sum to the size of the parent.
The random splitting rule is equivalent to randomly selecting a composition (with uniform distribution), so we term this model the random composition tree model.
In Fig.~\ref{fig:tree_numerics}a, we show a realization of a random composition tree, where the root has size $n=20$, and each node has $k=4$ children, some of which are empty.

The random composition tree model (with additional working memory capacity constraints) has been shown to predict the average length and degree of summarization in recall narratives of varying lengths \cite{Zhong2025}.
More recently, we showed that the model captures the chunk-size distributions of texts semantically chunked to the word level using a large language model (LLM) \cite{Zhong2026}.
By selecting the branching factor $k$ that yields the highest similarity to the random model, we found that different text types (children's stories, fiction, poetry) are characterized by different $k$'s (from 2 to 6). 
Crucially, these text types exhibit different entropy rates measured from LLM probabilities, and numerical calculations revealed that the entropy rate of the random trees follows an analogous trend with $k$.

Motivated by these findings, we set out to mathematically analyze the entropy of random composition trees and present three main results in this paper. First, we derive that the entropy grows linearly with the size of the root, as shown in Fig.~\ref{fig:tree_numerics}b.
Second, we define the entropy rate of the tree as the coefficient of the leading linear term, and calculate analytically how it depends on the branching factor $k$, as shown in Fig.~\ref{fig:tree_numerics}c.
Thirdly, we show that large trees exhibit typicality, meaning that with high probability, large trees belong to a typical set in which each tree is approximately equally likely.
Typicality can be demonstrated by the convergence (in probability) of the information rate -- the negative logarithm of the probability of occurrence of a particular tree, divided by the size of the root -- to its expected value, the entropy rate; see Fig.~\ref{fig:tree_numerics}d.
The mathematical analysis reveals that the properties of random composition trees parallel information-theoretic properties of language when treated as a stationary (ergodic) stochastic process.

\section{Entropy Recursion Equation}
In the random composition tree model, each node of size $n$ splits into $k$ possibly-empty chunks, with all splittings assumed to be equally likely. 
Each splitting constitutes a weak composition, an ordered sequence of $k$ non-negative integers, $\left( m_1, \dots, m_k \right)$, that sum to the parent's size, $n = \sum_{i=1}^{k} m_i$.
The number of possible splittings is given by the 'stars and bars' combinatorial expression ${n + k -1\choose k-1}$, which counts the number of ways to place $k-1$ divisions (bars) within $n$ elements (stars).
According to the model, the splitting continues recursively until a stopping condition is met: at a node with size $ m=0$ or $1$, or at a node with the same size as its parent, $m=n$.

We call the above formulation the 'original model' because it is the one we originally used to compare tree entropy rates against LLM entropy rates for texts with different corresponding branching factors $k$ \cite{Zhong2026}. 
Later in the paper, we analyze a 'resplitting' variant of the model, in which the $m=n$ stopping condition is removed, allowing a node to continue splitting even if it has the same size as its parent.
Both models are plausible descriptions of how a text can be recursively divided into coherent chunks.
In the original model, some chunks have sizes greater than 1 and cannot be split further.
In the resplitting variant, the resplitting events can be interpreted as delaying the division of a chunk into smaller parts to a deeper level.

We can obtain a recursive equation for the entropy, generalizing the derivation in \cite{Cover2005} (Problem 4.5) from binary to $k$-ary trees. To this end, we decompose the entropy $H(n)$ of a random tree with root size $n$ as a sum of two terms. The first term is the splitting entropy, $S(n)$, which is the entropy of the immediate splitting. For the random composition $k$-ary tree, each split is equally likely, so the entropy is simply the logarithm of the number of possible splits described above:
\begin{align}
S(n) = \log {n + k -1\choose k-1} 
\end{align}

The second term is the entropy of the resulting sub-trees, conditional on the immediate splitting.
It can be simplified by the fact that the sub-trees are independent and identically distributed, and by using the marginal probability $p_n(m)$ that a specific node has size $m$ when its parent has size $n$.
The probability $p_n(m)$ is the ratio of the number of splits that have a child of size $m$ to the total number of splits, which is given in our model by:
\begin{align}
p_{n}(m) &= \frac{{n-m + k -2\choose k-2}}{{n + k -1\choose k-1}}  \label{eq:p_knm} 
\end{align}

Combining, we get the following recursive equation for the entropy of a random composition $k$-ary tree:
\begin{align}
H(n) &= S(n) + k \sum_{m=2}^{n-1}p_{n}(m) H(m) \label{eq:H_recursive} 
\end{align}
The sum runs from $m=2$ to $n-1$ due to the aforementioned stopping conditions.

\section{Entropy Grows Linearly} \label{sec:entropy_linear}
In this section, we analyze how the entropy $H(n)$ scales with the size of the root $n$, for large $n$. 
We start with the 'resplitting' variant of the model, which is easier to solve. This amounts to the removal of the $m=n$ stopping condition, which turns equation \ref{eq:H_recursive} into the following equation:
\begin{align}
H_1(n) &= S(n) + k \sum_{m=2}^{n}p_{n}(m) H_1(m) \label{eq:H1_recursive} 
\end{align}

We can transform this discrete equation into a continuous equation using the Poissonization technique \cite{Szpankowski2001}, as follows.
It is easy to check that $p_{n}(m)$ from Eq.~\eqref{eq:p_knm} can be written as the following integral:
\begin{align}
p_n(m) &= \int_0^1 \binom{n}{m} p^m (1-p)^{n-m} f(p) dp \nonumber \\
f(p) &= (k-1)(1-p)^{k-2}
\label{eq:pnm1}
\end{align}
This representation can be understood as the result of a two-step process. First, we break the unit interval into $k$ parts by sampling $k-1$ division points uniformly (flat Dirichlet distribution), such that the length $p$ of each part has the same marginal probability density function $f(p)$. For each division, we sample $n$ points uniformly from the unit interval and count how many fall into each of the $k$ parts (a multinomial distribution), yielding the equation above. 

Next, insert Eq.~\eqref{eq:pnm1} into Eq.~\eqref{eq:H1_recursive} and apply the Poisson transform, $\tilde{f}(x) = \sum_{n=0}^\infty f(n) \frac{x^n e^{-x}}{n!}$, to both sides of the resulting equation. After changing the order of summation and simplifying, we get a continuous equation:
\begin{align}
\tilde{H}_1(x) = \tilde{S}(x) + k \int_0^1 f(p) \tilde{H}_1(px) dp
\end{align}
where $\tilde{H}_1(x)$ and $\tilde{S}(x)$ are, respectively, the entropy and the splitting entropy of a random composition tree where the root size is not fixed to $n$, but rather is a Poisson random variable with mean $x$.
The integral term in the continuous equation above results from the fact that, after Poissonization, the two-step process described above admits a simple recursive description. First, we break the unit interval into $k$ parts using the same distribution as before. Second, because the parent node size has a Poisson distribution with mean $x$, the number of points that fall into each part is also Poisson-distributed with mean $px$, where $p$ is the length of the part.

For smooth functions (like $x^r$ or $\log x$), $\tilde{f}(x)$ and $f(n)$ have the same leading-term asymptotic behavior for large $x$ and large $n$, respectively \cite{Szpankowski2001}, so we can analyze the asymptotic behavior of $\tilde{H}_1(x)$ to deduce the asymptotic behavior of $H_1(n)$.
To determine how $\tilde{H}_1(x)$ grows for large $x$, we apply the Mellin transform $\hat{F}(s)=\int_0^\infty F(x) x^{s-1} dx$ to both sides, use the scaling property of the Mellin transform, and solve for $\hat{H}_1(s)$:
\begin{align}
  \hat{H}_1(s) = \frac{\hat{S}(s)}{1-K(s)} \label{eq:H1_mellin}
\end{align}
where we defined the characteristic function:
\begin{align}
  K(s) = k \int_0^1 f(p) p^{-s} dp = \frac{k!}{\prod_{j=1}^{k-1} (j-s)} \label{eq:char_func}
\end{align}
The asymptotic behavior of $\tilde{H}_1(x)$ for large $x$ is determined by the poles of $\hat{H}_1(s)$: a simple pole at $s_0$ translates to $x^{-s_0}$ in the large $x$ asymptotic expansion of $\tilde{H}_1(x)$ \cite{Flajolet1995}.

The poles of $\hat{H}_1(s)$ are the poles of $\hat{S}(s)$ and the roots of the characteristic equation $K(s)=1$.
The characteristic equation $K(s)=1$ is well-studied in the analysis of algorithms literature, as it is the same one that arises in the analysis of $m$-ary random search trees \cite{Mahmoud1992}. 
For all $k$, we have $s=-1$ as a root of the characteristic equation, and all other roots are subdominant with real part larger than $-1$ (see Appendix \ref{app:char_eq}).

If $S(n)$ grows sublinearly with $n$ for large $n$, then $\hat{S}(s)$ has no poles with real part larger than $-1$, so that the dominant pole of $\hat{H}_1(s)$ is the root $s=-1$ of the characteristic equation, which means that $H_1(n)$ grows linearly with $n$ for large $n$.
Conversely, if $S(n)$ grows superlinearly with $n$, then $H_1(n)$ also grows superlinearly according to the leading term in the asymptotic expansion of $S(n)$.
In our case, $S(n)$ grows logarithmically with $n$, so $H_1(n)$ scales linearly for large $n$.

A similar analysis can be applied to the original model (with the $m=n$ stopping condition), where we modify the splitting entropy term: $S_{mod}(n) = S(n) - k p_n(n) H(n)$. We can rule out the possibility that $S_{mod}(n)$ grows faster than $n$, because in this case $H(n)$ would scale like $p_n(n) H(n)$, which is a contradiction because $p_n(n)$ goes like $n^{-k+1}$ for large $n$. Thus, the dominant pole of $\hat{H}(s)$ would be the same as before, and $H(n)$ also scales linearly for large $n$.

More generally, it can be proven that the entropy grows linearly with tree size for any stochastic $k$-ary tree model based on a recursive splitting process satisfying three properties.
First, each node splits independently, with child sizes distributed solely based on the parent's size. Second, each splitting satisfies node size conservation ($\sum_{i=1}^{k} m_i = n$). Third, for large parent size $n$, the expected number of size-$m$ children, $w_{n,m}$, can be approximated by a continuous scale-free shape function $\omega(z)$ with $z = m/n$, such that $w_{n, m} = \frac{1}{n}\omega\left(\frac{m}{n}\right) + O(n^{-2})$ (uniformly in $m$).
In our case, the expected number of children is $w_{n,m}=k p_n(m)$ and the shape function is $\omega(z) = k f(z)= k(k-1)(1-z)^{k-2}$.
We derive this more general result in Appendix \ref{app:linear_entropy}, using Roura's continuous master theorem \cite{Roura2001}.

\section{The Entropy Rate}
After we established that $H(n)$ scales linearly with $n$ for large $n$, we define the 'entropy rate' of the random composition tree ensemble as the coefficient of the leading linear term, in analogy with the entropy rate of stationary stochastic processes \cite{Cover2005}.

For the resplitting model variant, we can express the entropy rate $h_1$ via the negative residue of $\hat{H}_1(s)$ at the dominant pole $s=-1$:
\begin{align}
  h_1 = -\text{Res}_{s=-1} \hat{H}_1(s) = \frac{\hat{S}(-1)}{K'(-1)}
\end{align}
The numerator is given by $\hat{S}(-1) = \int_0^\infty \tilde{S}(x) x^{-2} dx$, where $\tilde{S}(x) = \sum_{n=0}^\infty S(n) \frac{x^n e^{-x}}{n!}$.
Substituting $S(0) = 0$ and $S(1) = 0$, changing the order of summation and integration, and evaluating the integral, we get $\hat{S}(-1)=\sum_{m=2}^\infty \frac{S(m)}{m(m-1)}$.
The denominator is given by $K'(-1)=\sum_{i=2}^k \frac{1}{i} = H_k - 1$, where $H_k$ is the $k$-th harmonic number.
The entropy rate is then given by:
\begin{align}
  h_1 = \frac{1}{H_k - 1} \sum_{m=2}^\infty \frac{S(m)}{m(m-1)}\label{eq:h1_rate}
\end{align}

For the original model, the modified splitting entropy term $S_{mod}(n)$ depends on $H(n)$, so we cannot obtain a closed-form expression for the entropy rate using the Mellin transform method.
We thus adopt a different approach based on the expected number of nodes of size $m$ in a tree of size $n$, denoted $A(n,m)$.
We can formulate a recursion equation for the expected number of nodes $A(n,m)$, with a form very similar to the entropy equation, and along the lines of the previous derivation, show that it scales linearly with $n$ for large $n$ (see Appendix \ref{app:a_m} for details).
In that case, we can write $A(n,m) \sim a_m n$ for large $n$, where $a_m$ is the (expected) growth rate of size $m$ nodes for large trees.

Using the growth rates $a_m$, the entropy rate is given by:
\begin{align}
h = \sum_{m=2}^{\infty} a_m S(m) \label{eq:h_rate}
\end{align}
This equation also applies to the resplitting model variant, so by comparing it with Eq.~\eqref{eq:h1_rate}, we can deduce that the growth rates are given by $\frac{1}{H_k - 1} \frac{1}{m(m-1)}$ for this variant, i.e. the growth rates are monotonically decreasing with $m$ and satisfy an $m^{-2}$ power law for large $m$, two properties that also hold for the original model, as we discuss below.

We can construct a 'backward' recursion equation for $a_m$, which relates the growth rate of size $m$ nodes to the growth rates of larger nodes:
\begin{align}
a_m = k \sum_{n=m+1}^{\infty}  p_{n}(m) a_n \label{eq:a_recursive} \,\, ,
\end{align}
from which one can show that the growth rates $a_m$ monotonically decrease with $m$ (see Appendix \ref{app:inverse_factorial}).
Eq.~\eqref{eq:a_recursive} is homogeneous, so to fully determine $a_m$ we use the following normalization condition that is based on node size conservation -- the root size $n$ is equal to the total size of all terminal nodes:
\begin{align}
  k \sum_{m=2}^{\infty} a_m (p_m(1) + m p_m(m)) = 1 \label{eq:a_norm}
\end{align}

In Appendix \ref{app:inverse_factorial}, we show that we can solve for $a_m$ by expanding using inverse factorial series:
\begin{align}
  a_m &= \mathcal{N} \sum_{j=2}^{\infty} \gamma_j \binom{m+j}{j}^{-1} 
  \label{eq:am}
\end{align}
where $\gamma_2=1$ and $\mathcal{N}$ is determined via the normalization condition. The coefficients $\gamma_j$ for $j \ge 3$ are determined recursively:
\begin{align}
  \gamma_i =  \frac{1}{1 - \beta_{i, i}} \sum_{j=2}^{i-1} \gamma_j \beta_{j, i} 
\end{align}
where the coefficients $\beta_{j, i}$ have closed-form expressions, given in Appendix \ref{app:inverse_factorial}. In Eq.~\eqref{eq:am}, the first term in the series dominates for large $m$, hence growth rates $a_m$ behave as a power law $a_m \sim m^{-2}$.

This solution simplifies considerably for the two cases. In the case of binary trees, we show in Appendix \ref{app:inverse_factorial} that $a_m = \frac{2}{(m+2)(m+3)}$, which can also be derived directly from the entropy recursion equation, as shown in Appendix \ref{app:binary_tree}.
In the case of a large splitting factor $k$, we show in Appendix \ref{app:inverse_factorial} that $a_m = \frac{1}{H_{k-1} - 1}\frac{1}{m(m-1)}$, which is very similar to the growth rates in the resplitting model.

\begin{figure}[htbp]
    \centering
    \includegraphics[width=0.4\textwidth]{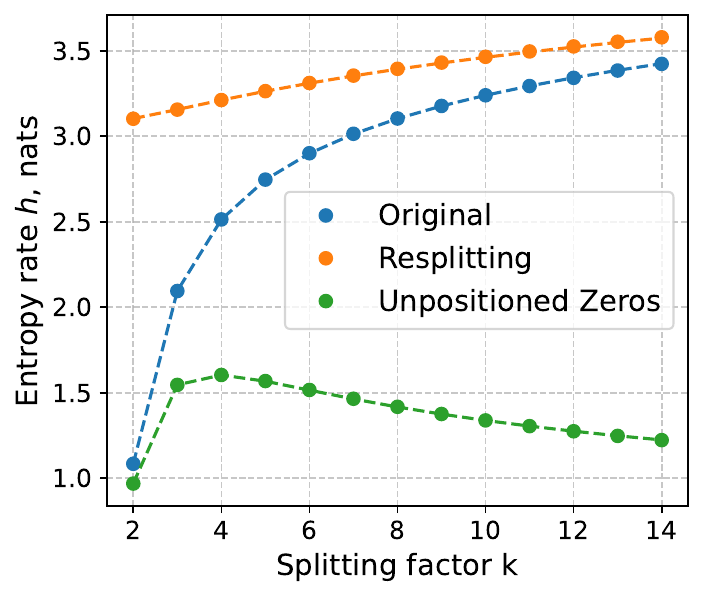}
    \caption{The entropy rate of the random composition tree model as a function of the splitting factor $k$: for the original model and two different variants. 
    In the resplitting model, when a node splits into a child of the same size as its parent, the child continues to split, while in the original model, this constitutes a stopping condition.
    In the 'unpositioned zeros' model, trees that differ only in the position of their empty chunks are considered equivalent.
    (Dashed lines are calculated analytically; points are calculated numerically with $n=2000$.)}
    \label{fig:entropy_comparison}
\end{figure}

While the $m=n$ stopping condition occurs with a negligible probability for nodes with large sizes, it changes considerably the growth rates of nodes with small sizes, and has a significant effect on the entropy rate, as can be seen in Figure \ref{fig:entropy_comparison}. For large $k$, the probability of a node to split into a child of the same size gets very small, also for small chunk sizes, so the entropy rates of the two models become similar.

In the original model and resplitting model variant, splittings that differ in the position of their empty chunks are considered distinct.
However, when splitting text into coherent chunks, the position of empty chunks between the non-empty ones does not have a clear interpretation, so we analyze another model variant, where we treat splittings that differ only in the position of their empty chunks as equivalent.
We thus group equivalent splittings into a single event with a probability that is larger by a multiplicity factor, which is the number of equivalent splittings (since all splittings are equally likely).
In this case, the splitting entropy can be decomposed into a sum over $i$, where $i$ is the number of non-empty chunks.
The probability of a specific group of equivalent splittings with $i$ non-empty chunks is ${k \choose i}/{n + k - 1 \choose k - 1}$.
Since there are ${n-1 \choose i-1}$ distinct splittings into $i$ non-empty chunks, we get the following expression for the splitting entropy of this model variant:
\begin{align}
S_{2}(n) = \log {n + k -1\choose k-1} - \sum_{i=1}^{\min(n, k)}\frac{ {n-1 \choose i-1} \cdot {k \choose i}}{{n + k - 1 \choose k - 1}} \log {k \choose i}
\end{align}
We then calculate the entropy rate of this model variant using Eq.~\eqref{eq:h_rate}, with the modified splitting entropy $S_{2}$ and the same growth rates $a_m$ of the original model.
As $S_2(n) < S(n)$, the entropy rate of this model variant is necessarily lower than that of the original model for all $k$s.
Interestingly, in this variant, the entropy rate as a function of the branching factor $k$ is non-monotonic and reaches its peak at $k=4$, as shown in Fig.~\ref{fig:entropy_comparison}.

\section{Typicality of Large Trees} \label{sec:typicality}
Up until now, we have analyzed the entropy of random composition trees, which is the expected information content of a tree.
The information content of a random composition tree $T_n$ of size $n$ is given by the negative logarithm of its probability to be realized under the random splitting process: $I(T_n) = -\log P(T_n)$.
In this section, we show that the information rate of the tree, which we define by $\frac{1}{n} I(T_n)$, converges in probability to the entropy rate $h$ for large $n$, a property we term typicality, in analogy with the typicality of ergodic stochastic processes \cite{Cover2005}.
Typicality means that for large $n$, there are about $e^{nh}$ typical trees, each is approximately equally probable with probability $e^{-nh}$, and the probability of observing non-typical trees decays to zero.

To derive typicality, we will analyze the variance of the information content, $V(n) = \text{Var}(I(T_n))$.
The variance can be calculated using the Law of Total Variance, by conditioning on the root split $\vec{M}$, which is the vector of sizes of the child nodes of the root:
\begin{align}
V(n) = \mathbb{E}[\text{Var}(I(T_n) \mid \vec{M})] + \text{Var}(\mathbb{E}[I(T_n) \mid \vec{M}])
\end{align}

Given the root partition $\vec{M}$, the sub-trees evolve independently, so the conditional variance is just the sum of the sub-tree variances:
$\text{Var}(I(T_n) \mid \vec{M}) = \sum_{i=1}^k \text{Var}(I(T_{M_i})) = \sum_{i=1}^k V(M_i)$.
Taking the expectation over all partitions gives the first term:
\begin{align}
  \mathbb{E}[\text{Var}(I(T_n) \mid \vec{M})]=\mathbb{E}\left[ \sum_{i=1}^k V(M_i) \right] = k \sum_{j=2}^{n-1} p_n(j) V(j)
\end{align}

For the second term, by linearity of expectation, the conditional expected log-likelihood is:
$\mathbb{E}[I(T_n) \mid \vec{M}] = \log p_n(\vec{M}) + \sum_{i=1}^k H(M_i)$.
Now, when we take the variance over all partitions, the first term vanishes as $p_n(\vec{M})$ is constant (all partitions are equally likely), and we are left with the variance of the sum of the sub-tree entropies:
\begin{align}
  U(n) = \text{Var}(\mathbb{E}[I(T_n) \mid \vec{M}]) = \text{Var}\left( \sum_{i=1}^k H(M_i) \right)
\end{align}
where the dependence on $n$ is due to the fact that $\sum_i M_i = n$.
In Appendix \ref{app:typicality}, we show that $U(n)$ is $O(n^{\beta})$ with $\beta < 2$ for large $n$.
The main idea is to use the linear scaling of the entropy, $H(n) = hn + r_n$, where $r_n$ is a sublinear term.
Using $\sum_i M_i = n$, we can rewrite the inhomogeneous term as a variance of a sum of sublinear terms: $U(n) = \text{Var}\left( \sum_{i=1}^k r_{M_i} \right)$, so that it has subquadratic growth with $n$.

Adding these two terms, we get the following recurrence for the variance of the information content:
\begin{align}
  V(n) = U(n) + k \sum_{j=2}^{n-1} p_n(j) V(j) \label{eq:V_recursive}
\end{align}
which has the same form as the entropy recurrence, Eq.~\eqref{eq:H_recursive}, but with a different inhomogeneous term.
In the recurrence analysis carried out in section \ref{sec:entropy_linear}, we established that if $U(n)$ scales sublinearly with $n$, then $V(n)$ scales linearly with $n$, and if $U(n)$ scales superlinearly with $n$, then $V(n)$ scales according to the leading term in the asymptotic expansion of $U(n)$.
As $U(n)$ is $O(n^{\beta})$ with $\beta < 2$, then so is $V(n)$, and therefore the variance of the information rate, $n^{-2} V(n)$, goes to zero for large $n$. 
This, in turn, implies that the information rate converges in probability to its expected value, the entropy rate $h$, establishing typicality.

The information rate can also be expressed as a sum over node sizes:
\begin{align}
\frac{1}{n} I(T_n) = \frac{S(n)}{n} + \sum_{m=2}^{n-1} \frac{N_m(T_n)}{n} S(m) \label{eq:info_rate}
\end{align}
where $N_m(T_n)$ is the number of nodes of size $m$ in a particular tree $T_n$ of size $n$.
We can do a similar variance analysis for $N_m(T_n)$, showing that the growth rate $\frac{1}{n}N_m(T_n)$ converges in probability to the expected growth rate $a_m$ for large $n$, with $m$ fixed (see Appendix \ref{app:typicality}). 
Moreover, based on Eq.~\eqref{eq:info_rate}, we can derive the convergence of the information rate to the entropy rate from the convergence of the growth rates to their expected values (see Appendix \ref{app:typicality}).
Thus, typical trees are characterized as having about $a_m n$ nodes of size $m$ for large $n$.

\section{Subextensive Entropy}
The entropy of a stationary stochastic process is extensive in the sequence length, i.e. $H(n)$ grows like $hn$ for large $n$. 
The subextensive part, $H(n)-hn$, is called predictive information as it is equal to the mutual information between a past sequence of length $n$ and an infinite future sequence that immediately follows \cite{Bialek2001}, and as such it is non-negative.
The excess entropy is defined as the limit of the predictive information for large $n$, $E = \lim_{n\to\infty} \left[H(n)-hn\right]$.
Based on a reanalysis of Shannon's estimate for the entropy of printed English, the 'relaxed Hilberg hypothesis' was proposed, which states that the subextensive part of the entropy of natural language grows as a power law $n^\alpha$ (with $\alpha <1$) \cite{Debowski2020}, which means that the excess entropy diverges for natural language.

In the random composition tree model, the subextensive part of the entropy can be found by analyzing the subdominant poles of $\hat{H}_1(s)$, given in Eq.~\eqref{eq:H1_mellin}.
There is a logarithmic subextensive term that is contributed from the double pole at $s=0$ of $\hat{S}(s)$, as the splitting entropy $S(n)$ grows logarithmically with $n$.
When we plug in $A \log n$ into the recurrence equation for $H(n)$, we get $A=-1$ for all $k$ (see Appendix \ref{app:subextensive}).
For small $k$, the subdominant poles contributed from the roots of the characteristic equation $K(s)=1$ all correspond to decaying subextensive terms.
Eventually, when $k$ is increased to 14, a pair of complex conjugate roots cross the imaginary axis in a Hopf bifurcation, which gives rise to a subextensive term that grows as a power-law with log-periodic oscillations (see Appendix \ref{app:subextensive}).

We thus conclude that the subextensive part of the entropy of random composition trees behaves very differently from the case of natural language treated as a stationary stochastic process. 
For $k\leq 13$, which covers the range of branching factors that we found to be relevant for natural language \cite{Zhong2026}, the subextensive part of the entropy is negative and grows logarithmically for large $n$, in contrast to the non-negativity of the predictive information of stationary stochastic processes, and in contrast to the relaxed Hilberg hypothesis.

\section{Discussion}
The random composition tree model is a statistical ensemble of trees that correspond to how texts are recursively split into coherent chunks, forming a hierarchical semantic structure.
In this paper, we analyzed key information-theoretic properties of this stochastic splitting model.
We showed that the entropy of the trees grows linearly with the root size, and derived an analytical expression for the entropy rate as a function of the branching factor $k$.
We showed that large trees exhibit typicality, i.e., the information rate converges in probability to the entropy rate.
Moreover, we analyzed the subextensive part of the entropy, and showed that its behavior is qualitatively different from that of natural language when modeled as a stationary stochastic process.

We also showed that the entropy rate is remarkably sensitive to 'microscopic' modifications - model changes that significantly affect small chunks but have a negligible impact on large ones.
We analyzed two model variants: the resplitting variant, which allows a child of the same size as its parent to split further, and the 'unpositioned zeros' variant, which treats splittings differing only in the position of empty chunks as equivalent.
The original model and the resplitting variant maintain a monotonically increasing entropy rate as the branching factor $k$ increases, matching the empirical trend observed in the entropy rates of texts with different branching factors \cite{Zhong2026}.
In contrast, the unpositioned zeros variant peaks at $k=4$ and decreases thereafter, demonstrating that this monotonic increasing trend is not a necessary consequence of recursive splitting.

The random composition tree model exhibits correspondence with two important informational properties of language when modeled as a stochastic process.
In the case of language as a stationary process, the entropy grows asymptotically linearly with the length of the text, and in the case of an ergodic process, the information rate converges to the entropy rate, i.e., exhibits typicality \cite{Cover2005}.
The random composition tree model shares these two properties despite being fundamentally different in nature.
In contrast to a discrete stochastic process that is defined as a sequence of random variables, each defined over a fixed set of possible outcomes (like an alphabet of words), our model is defined over a set of tree realizations that is distinct for every root size and expands combinatorially as the root size grows.
While the random composition tree model is stripped of any semantic or syntactic content, it may serve as a useful starting point for a model of language that incorporates the hierarchical semantic structure of texts, without the need to assume stationarity or ergodicity.

\vspace{+2em}
\section*{Acknowledgments}
W.Z. acknowledges funding from NSF ACCESS Discover Grant CIS240836, and support from the Simons foundation and the Eric and Wendy Schmidt Membership in Biology at the Insitute for Advanced Study. M.K. is supported in part by a grant from Fran Morris Rosman and Richard Rosman. M.T. is supported by the Simons Foundation, MBZUAI-WIS Joint Program for Artificial Intelligence Research and Foundation Adelis.

\newpage
{\printbibliography}

\newpage
\appendix

\makedivisiontitle{Appendices}
\section{The Characteristic Equation} \label{app:char_eq}
The characteristic equation is given by $K(s)=1$, where $K(s)$ is the characteristic function defined in Eq.~\eqref{eq:char_func}:
\begin{align}
  \prod_{j=1}^{k-1} (j-s) = k!
\end{align}
A characteristic equation root at $s_0$ translates to $n^{-s_0}$ in the large $n$ asymptotic expansion of the entropy $H(n)$.

A simple check shows that $s=-1$ is a solution.
We can show that all other roots are subdominant with real part larger than $-1$.
First, consider real roots. 
For $s<1$, the left-hand side is positive and strictly decreasing with $s$, so since $s=-1$ is a solution, it is the unique real root in this domain.
Second, consider complex roots $s = x + iy$ with $y \neq 0$. 
They must have $x=\mathrm{Re}(s)>-1$, otherwise we have $|j - s| > j+1$ for all $j \ge 1$ which implies $\prod_{j=1}^{k-1} |j - s| > \prod_{j=1}^{k-1} (j+1) = k!$, i.e. the characteristic equation is not satisfied.

\section{Conditions for Linear Entropy Growth} \label{app:linear_entropy}
In this appendix, we show that the linear scaling of the entropy holds generally for any stochastic $k$-ary tree model satisfying node size conservation, provided the transition probabilities admit a scale-free continuous limit, using Roura's continuous master theorem \cite{Roura2001}. 

We consider a general recursive splitting process in which a node of size $n$ splits into $k$ children with sizes $(m_1, \dots, m_k)$ satisfying size conservation:
\begin{align}
  \sum_{i=1}^{k} m_i = n \label{eq:conservation}
\end{align}
We assume the distribution of child sizes $(m_1, \dots, m_k)$ depends solely on the parent's size $n$.
Under this assumption, we can derive a recursive equation for the entropy $H(n)$ of a tree of size $n$, along the lines of how Eq.~\eqref{eq:H_recursive} was derived in section \ref{sec:entropy_linear}:
\begin{align}
  H(n) = S(n) + \sum_{m=1}^{n-1} w_{n,m} H(m) \label{eq:H_general_rec}
\end{align}
where $w_{n,m}$ is the expected number of children of size $m$ generated by a parent of size $n$ and $S(n)$ is the splitting entropy.
Note that we do not assume that the marginal distribution of a single child's size is the same for all children (i.e., the children are not necessarily exchangeable). Moreover, we do not assume that all the splittings are equally likely as in the random composition tree model.
In Eq.~\eqref{eq:H_general_rec}, we set $H(0)=0$ and assume that $w_{n,n}=0$, so we sum over $1 \leq m < n$ (if a parent can produce a single child of the same size with probability $w_{n,n} < 1$, we can simply subtract $w_{n,n}H(n)$ from both sides and renormalize by $1 - w_{n,n}$ to get a valid recurrence equation with $\widetilde{S}(n) = S(n) / (1 - w_{n,n})$ and $\widetilde{w}_{n,m} = w_{n,m} / (1 - w_{n,n})$ for $m < n$).

We assume that for large $n$, the expected child count $w_{n,m}$ is well-approximated by a continuous shape function $\omega(z)$ (with continuous derivative) over the unit interval $z \in (0,1)$:
\begin{align}
  w_{n,m} = \frac{1}{n} \omega\left(\frac{m}{n}\right) + O(n^{-2}) \label{eq:shape_function_def}
\end{align}
uniformly in $1\leq m < n$ (this is a stronger assumption that implies the definition of the shape function in \cite{Roura2001}).
In the random composition tree model, $w_{n,m} = k p_n(m)$, where $p_n(m)$ is the marginal distribution of a single child's size, and $\omega(z) = k f(z) = k(k-1)(1-z)^{k-2}$.

Recurrences of the form \eqref{eq:H_general_rec} with shape functions satisfying \eqref{eq:shape_function_def} can be analyzed asymptotically using Roura's continuous master theorem \cite{Roura2001}. 
Specifically, if $\int_0^1 z \, \omega(z)  \,dz=1$ and the inhomogeneous term $S(n)$ is sublinear, then the solution of the recurrence scales linearly with $n$ for large $n$.
Indeed, due to node size conservation, we have $\sum_{m=1}^{n-1} m \, w_{n,m} = n$.
Using Eq.~\eqref{eq:shape_function_def}, and taking the large $n$ limit so that the sum can be replaced by an integral, we get: $\int_0^1 z \, \omega(z)  \, dz = 1$.
Additionally, the splitting entropy $S(n)$ is bounded by the entropy of the uniform distribution over all possible splittings, which is $O(\log n)$, so it is sublinear.

The above result can be extended to the case where node size conservation is satisfied only asymptotically for large $n$, as the size losses disappear in the large $n$ limit.
Such a case arises for random $k$-ary search trees, which store a set of ordered items (such as numbers) in a hierarchical tree structure, under the assumption that items are inserted in a random order (uniform distribution over all permutations).
Each node contains $k-1$ sorted values that divide the range of possible items into $k$ non-overlapping intervals, directing any further items into the corresponding $k$ child subtrees.
By defining node size as the total number of items in the subtree rooted at that node, the relation between the parent size and the sizes of the child nodes is given by $\sum_{i=1}^{k} m_i = n^\prime$, where $n^\prime = n - k + 1$, which satisfies size conservation asymptotically for large $n$.
Because all non-negative partitions of the remaining items are equally likely under a random insertion order, the expected number of subtrees of size $m$ is given by $w_{n,m} = k p_{n'}(m)$, where we use the marginal node size distribution of the random composition tree model with $n$ replaced by $n^\prime$. Since $n^\prime/n \to 1$ for large $n$, this yields the same continuous shape function as in the random composition tree model, $\omega(z) = k(k-1)(1-z)^{k-2}$.
Therefore, the entropy of random $k$-ary search trees also scales linearly with the root size for large $n$, as shown in \cite{Golkebiewski2018}, which also includes a closed-form expression for the entropy rate.

\section{Growth Rate of Nodes with Different Sizes} \label{app:a_m}
We define $A(n,m)$ to be the expected number of nodes of size $m$ in a tree of size $n$. Then we can write the entropy as:
\begin{align}
H(n) = S(n) + \sum_{m=2}^{n-1} A(n,m) S(m) 
\end{align}

We can formulate a recursion equation for the expected number of nodes $A(n,m)$:
\begin{align}
A(n,m) = k p_{n}(m) + k \sum_{j=m+1}^{n-1} p_{n}(j) A(j,m)  \label{eq:A_recursive}
\end{align}
This equation has an intuitive interpretation: the expected number of descendants of size $m$ for a node of size $n$ is equal to the expected number of direct children of size $m$ (the first term) plus the expected number of descendants of size $m$ that arise from the sub-trees of the direct children (the second term).

By defining $A(j,m) = 0 $ for $j \leq m$, the recursion equation for $A(n,m)$ has the same form as the entropy recursion equation, Eq.~\eqref{eq:H_recursive}, only with a different inhomogeneous term.
As the inhomogeneous term is sublinear in $n$ for large $n$ (for fixed $m$, $p_n(m)$ goes like $n^{-1}$ for large $n$), we can use the same analysis as in section \ref{sec:entropy_linear} to show that $A(n,m)$ also scales linearly with $n$ for large $n$.

In that case, we can write $A(n,m) \sim a_m n$ for large $n$, and the entropy rate is given by:
\begin{align}
h = \sum_{m=2}^{\infty} a_m S(m) \label{eq:h_rate1}
\end{align}

We can also formulate a 'backward' recurrence for $A(n,m)$:
\begin{align}
A(n,m) = k p_{n}(m) + k \sum_{j=m+1}^{n-1}  p_{j}(m) A(n,j) \label{eq:A_recursive_back}
\end{align}
The second term is again the expected number of non-direct descendants of size $m$, but now we sum over the parent size of the size-$m$ nodes (rather than over the size of the direct children of the root node).
We call this a 'backward' recurrence as we sum over larger problems as opposed to the regular 'forward' recurrence, where we sum over smaller problems.

By plugging $A(n,m) = a_m n$ into Eq.~\eqref{eq:A_recursive_back}, and taking large $n$ limit ($\frac{k}{n} p_{n}(m)\rightarrow 0$), we get a 'backward' equation for $a_m$:
\begin{align}
a_m = k \sum_{n=m+1}^{\infty}  p_{n}(m) a_n \label{eq:a_m_recursive}
\end{align}

To derive a normalization condition, we use the node size conservation -- $n$ is equal to the total size of all terminal nodes, and in particular to its expected value:
\begin{align}
  n = k(p_n(1) + n p_n(n)) + k \sum_{m=2}^{n-1} A(n,m) (p_m(1) + m p_m(m)) 
\end{align}
Taking large $n$ limit and using $A(n,m) = a_m n$, we get:
\begin{align}
  k \sum_{m=2}^{\infty} a_m (p_m(1) + m p_m(m)) = 1  \label{eq:a_m_norm}
\end{align}

The above relations - Eqs. \eqref{eq:h_rate1}, \eqref{eq:a_m_recursive}, and \eqref{eq:a_m_norm} - rely on interchanging the large $n$ limit and the summation. 
These interchanges can be justified by Tannery's theorem, which states that $\lim_{n\to\infty} \sum_{k=0}^n a_k(n) = \sum_{k=0}^{\infty} \lim_{n\to\infty} a_k(n)$, if $|a_k(n)| \leq M_k$ and $\sum_{k=0}^{\infty} M_k$ converges.
Such $n$-independent bounding sequences with a convergent series can be constructed for all three cases, using the bound $A(n,m) \leq \frac{2n}{m(m+1)}$, which can be established by induction from Eq.~\eqref{eq:A_recursive}.

\section{Solution using Inverse Factorial Series} \label{app:inverse_factorial}

In this appendix, we will show that the backward recurrence equation for $a_m$ can be solved using an inverse factorial series expansion.
Inverse factorials are functions of the following form:
\begin{align}
  \phi_r(n) = \binom{n+r}{r}^{-1} = \frac{r!}{(n+1)(n+2)\cdots(n+r)}
\end{align}

\subsection{Inverse Factorial Series}
The backward recurrence equation for $a_m$ in Eq.~\eqref{eq:a_recursive} can be written as $\mathcal{S}[a](m)=a_m$, where $\mathcal{S}$ is the summation operator defined by:
\begin{align}
  \mathcal{S}[a](m) = k \sum_{n=m+1}^{\infty} p_n(m) a_n
\end{align}

In subsection \ref{sec:complete_monotone} we show that $a_m$ is completely monotone, i.e. $(-1)^j \Delta^j a_m \ge 0$ for all $m \ge 0$ and $j \ge 0$, where $\Delta$ is the forward difference operator: $\Delta a_m = a_{m+1} - a_m$.
Using the Hausdorff moment theorem \cite{Feller1971}, we can then write $a_m$ as a sequence of moments of a unique probability distribution on the unit interval:
\begin{align}
  a_m = \int_0^1 z^m W(z) \, dz
\end{align}

Then, in subsection \ref{sec:continuous_equation}, we derive a continuous equation for $W(z)$, $\mathcal{V}[W](z) =W(z)$, where $\mathcal{V}$ is the following integral operator:
\begin{align}
  \mathcal{V}[W](z) = k(k-1) \left[ (1-z)^{-(k-1)} - 1 \right] \int_z^1 \frac{dy}{y} W(y) \left(1-\frac{z}{y}\right)^{k-2}
\end{align}

In subsection \ref{sec:coeffs}, we show that the integral operator $\mathcal{V}$ acts in an upper-triangular manner on the following basis of polynomial functions, $w_r(y) = r(1-y)^{r-1}$.
The moment sequence corresponding to these basis functions is given by inverse factorials, as $\int_0^1 z^m w_r(z) dz = \phi_r(m)$.
This means that the discrete summation operator is also upper-triangular, over inverse factorials:
\begin{align}
  \mathcal{S}[\phi_r](m) = \sum_{j=0}^{\infty} \beta_{r, r+j} \, \phi_{r+j}(m) \label{eq:upper_triang}
\end{align}
The coefficients $\beta_{r, r+j}$ are calculated explicitly in subsection \ref{sec:coeffs} and are given by:
\begin{align}
  \beta_{r, r+j} &= \frac{r \lambda_r}{r+j} \left[ H_j - \mathbb{I}_{j \ge k-1} H_{j-k+1} \right] \\
  \lambda_r &= \frac{k!}{(r)_{k-1}} \\
  H_n &= \frac{(r)_n (k-1)_n}{(r+k-1)_n n!}
\end{align}
where $(x)_n = x(x+1)\cdots(x+n-1)$ is the rising factorial, and $\mathbb{I}_{j \ge k-1}$ is the indicator function that is equal to $1$ if $j \ge k-1$ and $0$ otherwise.

We then plug an inverse factorial series expansion, $a_m = \sum \gamma_j \phi_{j}(m)$, into the fixed-point equation $a_m = \mathcal{S}[a](m)$ and match coefficients for each basis function. The first non-zero coefficient $\gamma_j$ in the expansion requires that $\beta_{j, j} = 1$. Indeed, $\beta_{2,2}=1$, which corresponds to a dominant power-law behavior of $a_m \sim m^{-2}$ for large $m$.
This leads us to the following inverse factorial series:
\begin{align}
  a_m &= \mathcal{N} \sum_{j=2}^{\infty} \gamma_j \phi_{j}(m) \label{eq:a_m_expansion}
\end{align}
We define $\gamma_2=1$ and determine $\mathcal{N}$ via the normalization condition, Eq.~\eqref{eq:a_norm}:
\begin{align}
\mathcal{N} =  k \sum_{j=2}^{\infty} \gamma_j \sum_{m=2}^{\infty} \phi_{j}(m) (p_m(1) + m p_m(m)) \label{eq:norm_factor}
\end{align}
By matching coefficients for each basis function, we obtain for $i \geq 3$:
\begin{align}
  \gamma_i =  \frac{1}{1 - \beta_{i, i}} \sum_{j=2}^{i-1} \gamma_j \beta_{j, i} \label{eq:gamma_coeff}
\end{align}

\subsection{Binary Tree}
In the case of $k=2$, the coefficients simplify to: 
\begin{align}
\beta_{r, r+p} = \frac{-2r}{(r+p)^2 (r+p-1)}
\end{align}
Using Eq.~\eqref{eq:gamma_coeff}, we get $\gamma_3=-2/3$ and all other coefficients are zero, so the unnormalized solution is given by:
\begin{align}
a_m &= \gamma_0 \phi_2(m) + \gamma_1 \phi_3(m) \nonumber = \frac{2}{(m+1)(m+2)} \left[ 1 - \frac{2}{m+3} \right] = \frac{2}{(m+2)(m+3)}
\end{align}
For this $a_m$, the normalization coefficient $\mathcal{N}$ turns out to be $1$.

\subsection{Large Splitting Factor}
To obtain a solution for large splitting factors, we calculate the coefficients in the limit $k \to \infty$. In this limit, the eigenvalues $\lambda_r = k! / (r)_{k-1}$ decay rapidly to zero for all $r \ge 3$, while $\lambda_2 = 1$.
For large $k$, we have $\lambda_i \to 0$ for $i \ge 3$, and correspondingly only the coefficients $\beta_{2, i}$ remain non-zero such that: $\gamma_i = \beta_{2, i}$. As $k \to \infty$, we have $\beta_{2, i} \to 2(i-1)/i$ and so the inverse factorial series expansion of $a_m$ becomes:
\begin{align}
  a_m = \mathcal{N} \sum_{i=2}^{\infty} \frac{2(i-1)}{i} \binom{m+i}{i}^{-1}
\end{align}

We can simplify this expression by splitting the summand into two terms:
\begin{align}
  a_m = 2\mathcal{N} m! \sum_{i=2}^{\infty} \left[ \frac{i!}{(m+i)!} - \frac{(i-1)!}{(m+i)!} \right] = 2\mathcal{N} m!  \left[\sum_{n=0}^{\infty} \frac{(n+2)!}{(m+n+2)!} - \sum_{n=0}^{\infty} \frac{(n+1)!}{(m+n+2)!}\right] 
\end{align}
Using the following relation: $\sum_{n=0}^{\infty} \frac{(n+c)!}{(m+n+c)!} = \frac{1}{m-1} \frac{c!}{(m+c-1)!}$ (which can be derived by rewriting the sum as a telescoping series), and after simplifying, we obtain:
\begin{align}
  a_m = \frac{2\mathcal{N}}{m(m-1)}
\end{align}

By plugging this expression into the normalization condition, Eq.~\eqref{eq:a_norm}, and evaluating the sums, we get:
\begin{align}
  \mathcal{N} = \frac{1}{2} \left( H_{k-1} - 1 + \frac{1}{k-1} + \frac{2}{k} \right)
\end{align}
This means that for large $k$, we have $a_m = \frac{1}{H_{k-1} - 1}\frac{1}{m(m-1)}$.

\subsection{Complete Monotonicity} \label{sec:complete_monotone}
We want to show that $a_m$ is a completely monotone sequence, i.e., $(-1)^j \Delta^j a_m \ge 0$ for all $m \ge 0$ and $j \ge 0$ (for any integer $k\geq 2$).
Here we will derive complete monotonicity not only for $a_m$, but for a family of $k-1$ sums for $i = 0, 1, \dots, k-2$:
\begin{align}
S_i(m) = k \sum_{n=m+1+i}^{\infty} \frac{\binom{n-m+k-2-i}{k-2-i}}{\binom{n+k-1}{k-1}} a_n
\end{align}
Notice that $S_0(m) = a_m$ via the backward recurrence.

We perform a simultaneous strong induction on $j$ across all $S_i(m)$.
Our base case ($j=0$): $S_0(m) \geq 0$, is satisfied for all $i$ since $a_n \ge 0$.
We now assume the inductive hypothesis, $(-1)^p \Delta^p S_i(m) \ge 0$, holds for all $p \le j$ and all $i$, and we will show that it holds for $p = j+1$ as well.

To do so, we express $-\Delta S_i(m) = S_i(m) - S_i(m+1)$ in terms of $S_{i+1}(m)$:
\begin{align}
-\Delta S_i(m) &= c_{i,m} a_{m+1+i} + S_{i+1}(m) \\
c_{i,m} &= k \frac{\binom{k-1}{k-2-i}}{\binom{m+i+k}{k-1}}
\end{align}
For $i=k-2$, only the first term exists, so we define $S_{k-1}(m) = 0$. As a sequence of $m$, $c_{i,m}$ is completely monotone, as can be shown by direct calculation.

We then apply $(-1)^j \Delta^j$ to this expression and obtain:
\begin{align}
(-1)^{j+1} \Delta^{j+1} S_i(m) = (-1)^j \Delta^j \left( c_{i,m} a_{m+1+i} \right) + (-1)^j \Delta^j S_{i+1}(m)
\end{align}

By the hypothesis, $a_{m+1+i}=S_0(m+1+i)$ is completely monotone up to order $j$. $c_{i,m}$ is also completely monotone. The product of $c_{i,m}$ and $a_{m+1+i}$ is also completely monotone up to order $j$, as can be shown using the discrete product rule: $\Delta^j (u_m v_m) = \sum_{i=0}^j \binom{j}{i} (\Delta^{j-i} u_{m+i}) (\Delta^i v_m)$. This makes the first term positive, while the second term is positive by the hypothesis. Therefore, $(-1)^{j+1} \Delta^{j+1} S_i(m) \ge 0$ for all $m \ge 0$ and $i = 0, 1, \dots, k-2$, which completes the induction.

\subsection{Derivation of the Continuous Equation} \label{sec:continuous_equation}
We substitute $a_n = \int_0^1 y^n W(y) \, dy$ directly into the backward recurrence:
\begin{align}
  a_m = k \sum_{n=m+1}^{\infty} p_n(m) \int_0^1 y^n W(y) \, dy = \int_0^1 dy \, W(y) \left[ k \sum_{n=m+1}^{\infty} p_n(m) y^n \right]
\end{align}

We can evaluate the inner sum by applying the integral representation of the inverse binomial coefficient, $\binom{n+k-1}{k-1}^{-1} = (k-1) \int_0^1 x^n (1-x)^{k-2} \, dx$, to the splitting probability:
\begin{align}
  p_n(m) = \binom{n-m+k-2}{k-2} (k-1) \int_0^1 x^n (1-x)^{k-2} \, dx
\end{align}
Substituting this into the summation and swapping the order of the sum and integral yields:
\begin{align}
  k \sum_{n=m+1}^{\infty} p_n(m) y^n = k(k-1) \int_0^1 dx \, (1-x)^{k-2} \sum_{n=m+1}^{\infty} \binom{n-m+k-2}{k-2} (xy)^n
\end{align}

We rewrite the series, by a  shift of index, using $t = n - m$. This factors out $(xy)^m$:
\begin{align}
  \sum_{t=1}^{\infty} \binom{t+k-2}{k-2} (xy)^{m+t} = (xy)^m \sum_{t=1}^{\infty} \binom{t+k-2}{k-2} (xy)^t
\end{align}
Recognizing the series expansion for the negative binomial, $(1-xy)^{-(k-1)} = \sum_{t=0}^{\infty} \binom{t+k-2}{k-2} (xy)^t$, we subtract the $t=0$ term (which is $1$) to close the sum:
\begin{align}
  \sum_{t=1}^{\infty} \binom{t+k-2}{k-2} (xy)^t = (1-xy)^{-(k-1)} - 1
\end{align}
Substituting this back gives the continuous convolution form over the scale-free splitting kernel:
\begin{align}
  k \sum_{n=m+1}^{\infty} p_n(m) y^n = k(k-1) \int_0^1 dx \, (xy)^m \left[ (1-xy)^{-(k-1)} - 1 \right] (1-x)^{k-2}
\end{align}

We substitute this expression back into the integral for $a_m$, and change variables in the inner integral by letting $z = xy$:
\begin{align}
  a_m = \int_0^1 dy \, W(y) k(k-1) \int_0^y \frac{dz}{y} \, z^m \left[ (1-z)^{-(k-1)} - 1 \right] \left(1-\frac{z}{y}\right)^{k-2}
\end{align}
Finally, changing the order of integration yields:
\begin{align}
  a_m = \int_0^1 dz \, z^m \left[ k(k-1) \left( (1-z)^{-(k-1)} - 1 \right) \int_z^1 \frac{dy}{y} W(y) \left(1-\frac{z}{y}\right)^{k-2} \right]
\end{align}

Since $a_m = \int_0^1 z^m W(z) \, dz$ for all $m$, we can equate the integrands to obtain the continuous equation for $W(z)$:
\begin{align}
  W(z) = k(k-1) \left[ (1-z)^{-(k-1)} - 1 \right] \int_z^1 \frac{dy}{y} W(y) \left(1-\frac{z}{y}\right)^{k-2}
\end{align}
We write this in operator form as $\mathcal{V}[W](z) = W(z)$.

\subsection{Derivation of the Coefficients} \label{sec:coeffs}
We evaluate the action of the operator $\mathcal{V}$ on a single basis function $w_r(y) = r(1-y)^{r-1}$:
\begin{align}
  \mathcal{V}[w_r](z) = k(k-1)r \left[ (1-z)^{-(k-1)} - 1 \right] \int_z^1 \frac{dy}{y} (1-y)^{r-1} \left(1-\frac{z}{y}\right)^{k-2}
\end{align}

Let $I(z)$ denote the inner integral: $I(z) = \int_z^1 y^{1-k} (1-y)^{r-1} (y-z)^{k-2} dy$.
We define $x = 1-z$, and apply the substitution $y = 1 - xu$:
\begin{align}
    I(z) &= x^{r+k-2} \int_0^1 (1-xu)^{-(k-1)} u^{r-1} (1-u)^{k-2} du
\end{align}

Next, we expand $(1-xu)^{-(k-1)} = \sum_{n=0}^{\infty} \frac{(k-1)_n}{n!} (xu)^n$, using the generalized binomial series, where $(x)_n = x(x+1)\cdots(x+n-1)$ is the rising factorial:
\begin{align}
    I(z) &= x^{r+k-2} \sum_{n=0}^{\infty} \frac{(k-1)_n}{n!} x^n \int_0^1 u^{r+n-1} (1-u)^{k-2} du
\end{align}
The integral equals to a Beta function, $\int_0^1 u^{r+n-1} (1-u)^{k-2} du = \frac{(k-2)!}{(r+n)_{k-1}}$, which gives:
\begin{align}
    I(z) &= x^{r+k-2} \sum_{n=0}^{\infty} \frac{(k-1)_n}{n!} \frac{(k-2)!}{(r+n)_{k-1}} x^n
\end{align}

We now substitute $I(z)$ back into the full operator definition $\mathcal{V}[w_r](z)$ and simplify:
\begin{align}
    \mathcal{V}[w_r](z) &= r \frac{k!}{(r)_{k-1}} \left[ x^{r-1} - x^{r+k-2} \right] \sum_{n=0}^{\infty} \frac{(r)_n (k-1)_n}{(r+k-1)_n n!} x^n
\end{align}

To express this in terms of the basis functions, we use $w_{r+j} = (r+j) x^{r+j-1}$, and so we arrive at an upper-triangular expansion of the operator $\mathcal{V}$ over the basis functions:
\begin{align}
  \mathcal{V}[w_r](z) = \sum_{j=0}^{\infty} \beta_{r, r+j} \, w_{r+j}(z)
\end{align}
where:
\begin{align}
  \beta_{r, r+j} &= \frac{r \lambda_r}{r+j} \left[ H_j - \mathbb{I}_{j \ge k-1} H_{j-k+1} \right] \\
  \lambda_r &= \frac{k!}{(r)_{k-1}} \\
  H_n &= \frac{(r)_n (k-1)_n}{(r+k-1)_n n!}
\end{align}

\subsection{Numerical Convergence and Truncation}
\begin{figure}[htbp]
    \centering
    \includegraphics[width=0.48\textwidth]{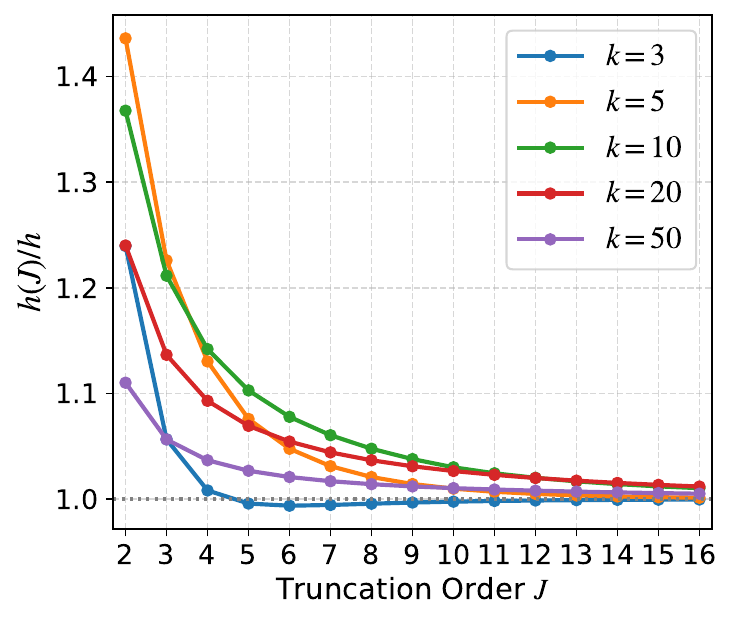}
    \caption{Convergence of the normalized entropy rate $h(J)/h$ as a function of the truncation order $J$ across different values of $k$.}
    \label{fig:inverse_factorial_convergence}
\end{figure}

Finally, we show the numerical convergence of the entropy rate in Fig.~\ref{fig:inverse_factorial_convergence} when truncating the expansion over $j$ up to an upper cutoff order $J$. In this truncation, the upper summation index is replaced by $J$ in both the expansion coefficients $a_m$, Eq.~\eqref{eq:a_m_expansion}, and the normalization constant $\mathcal{N}$, Eq.~\eqref{eq:norm_factor}. To ensure that truncation error in the outer series of Eq.~\eqref{eq:h_rate1} is negligible compared to the effects of $J$, we include sufficiently many terms in the $m$-summation. Specifically, because the summand decays slowly as $\sim \ln(m)/m^2$, yielding a truncation error scaling as $\sim \ln(m)/m$, we evaluate it up to $m=5000$. The resulting estimate $h(J)$, computed according to Eq.~\eqref{eq:h_rate} and normalized by the entropy rate value $h$, is shown as a function of the truncation order $J$ across different values of $k$. For small cutoff orders, the truncated sum overestimates $h$, while keeping terms up to $J \ge 10$ provides an accurate approximation across all evaluated values of $k$.

\section{Binary Tree Solution} \label{app:binary_tree}
\subsection{Original Model}
For the binary tree case, $k=2$ and $p_n(m)=\frac{1}{n+1}$, so the recursion relation becomes:
\begin{align}
H(n) = S(n) + \frac{2}{n+1} \sum_{m=2}^{n-1} H(m)
\end{align}

This recurrence relation has the same form as the one governing the average number of comparisons in the ``quicksort'' array sorting algorithm, which can be solved exactly using a differencing trick, as shown by Knuth \cite{Knuth1998}. Cover and Thomas were the first to apply the same technique to calculate the entropy of random binary trees, given as an exercise in their seminal book on information theory (Problem 4.5 in \cite{Cover2005}). While their exercise corresponds to a random strong integer-ordered partition with $k=2$, we apply it to our weak ordered-partition random tree model with $k=2$. Despite this difference in the underlying combinatorics, the form of the recurrence remains the same, and thus can be solved using the same differencing trick.

The first step is to get rid of the summation term by considering the difference $(n+1) H(n) - n H(n-1)$, which leads to an exactly solvable first-order linear recurrence relation. The solution is given by:
\begin{align}
H(n) = S(n) + 2(n+2) \sum_{m=2}^{n-1} \frac{S(m)}{(m+2)(m+3)}
\end{align}
So the entropy rate $C$ is given by:
\begin{align}
C = 2 \sum_{m=2}^{\infty} \frac{S(m)}{(m+2)(m+3)}
\end{align}

\subsection{Resplitting Model Variant}
For the resplitting model variant, we remove the $m=n$ stopping condition, resulting in the following equation:
\begin{align}
H(n) = S(n) + \frac{2}{n+1} \sum_{m=2}^{n} H(m)
\end{align}

Doing the same differencing trick, we can solve a similar first-order linear recurrence relation, and we arrive at the following solution:
\begin{align}
  H(n) = \frac{n+1}{n-1}S(n) + 2n \sum_{m=2}^{n-1} \frac{S(m)}{m(m-1)}
\end{align}
The entropy rate $C$ is then given by:
\begin{align}
C = 2 \sum_{m=2}^{\infty} \frac{S(m)}{m(m-1)}
\end{align}

\section{Typicality} \label{app:typicality}
\subsection{Bounding the Inhomogeneous Term} \label{app:inhom_bound}
Here we will show that the inhomogeneous term, $U(n)$, in the variance recurrence Eq.~\eqref{eq:V_recursive} is bounded by $O(n^{\beta})$ with $\beta < 2$ for large $n$.

In the main text, we established: $U(n) = \text{Var}\left( \sum_{i=1}^k H(M_i) \right)$, but to be more precise, for nodes of size $M_i$ that satisfy the stopping conditions, we have no contribution to the entropy, so we should write:
\begin{align}
  U(n) = \text{Var}\left( \sum_{i=1}^k H(M_i) \mathbb{I}_{2\leq M_i \leq n} \right)
\end{align}

We have shown that the entropy scales linearly, so that $H(n) = hn + r_n$, where $r_n$ is a sublinear term. by defining $r_0=0$, $r_1=-h$, and $r_n=-hn$ for the cases with stopping conditions, we can rewrite $\sum_{i=1}^k H(M_i) \mathbb{I}_{2\leq M_i \leq n} = h n + \sum_{i=1}^k r_{M_i}$, and so  $U(n) = \text{Var}\left( \sum_{i=1}^k r_{M_i} \right)$.

For ease of notation we denote $R=\sum_{i=1}^k r_{M_i}$, and we bound the variance by its second moment: $U(n) = \text{Var}(R) \le \mathbb{E}[R^2]$.
We then split the second moment into two complementary cases:
\begin{align}
  U(n) \le \mathbb{E}\left[R^2 \mathbb{I}_{M_i\ne n}\right] + \mathbb{E}\left[ (hn)^2 \mathbb{I}_{M_i=n}\right]
\end{align}

The first term can be bounded using the fact that $|r_n|$ is sublinear, $O(n^\alpha)$ with $\alpha < 1$ for large $n$, when $M_i\ne n$.
We have:
\begin{align}
\vert R \vert = \left\vert \sum_{i=1}^k r_{M_i} \right\vert \le \sum_{i=1}^k \vert r_{M_i}\vert \le C \sum_{i=1}^k M_i^\alpha
\end{align}
Then, using Jensen's inequality for the concave function $x^\alpha$, we can bound the sum of the $M_i^\alpha$ terms:
\begin{align}
\frac{1}{k}\sum_{i=1}^k M_i^\alpha \le \left( \frac{1}{k}\sum_{i=1}^k M_i \right)^\alpha = \left( \frac{n}{k} \right)^\alpha
\end{align}
Therefore $\vert R \vert$ is $O(n^\alpha)$, and so the first term is bounded by $O(n^{2\alpha})$.

The second term is calculated using the probability of the event that $M_i=n$ for some $i$: $k \cdot p_n(n) = \frac{k}{\binom{n+k-1}{k-1}}$, which is $O(n^{-(k-1)})$.
Therefore, the second term is bounded by $O(n^2) \cdot O(n^{-(k-1)}) = O(n^{3-k})$, which is $O(n)$ for $k=2$ and $O(1)$ for $k \ge 3$.
All together, we have $U(n) = O(n^{\beta})$ with $\beta = \max(2\alpha, 3-k)$, which is less than $2$ for all $k \ge 2$.

\subsection{Typicality of the Growth Rates}
We will show that the growth rate $\frac{1}{n}N_m(T_n)$ converges in probability to the expected growth rate $a_m$ for large $n$, by performing a variance analysis for $N_m(T_n)$, similar to the variance analysis for the information content in section \ref{sec:typicality}.

We use the Law of Total Variance, rewrite the two terms using $V(n) = \text{Var}(N_m(T_n))$ 
and $A(n,m)=\mathbb{E}[N_m(T_n)]$, and get a recurrence relation that has the same form:
\begin{align}
  V(n) = U(n) + k \sum_{j=2}^{n-1} p_n(j) V(j) 
\end{align}
We note that $V(j)=0$ for $j \le m$ since $N_m(T_n)$ is 0 for $n<m$ and 1 for $n=m$.
The only difference is in the inhomogeneous term $U(n)$, now given by:
\begin{align}
  U(n) = \text{Var}\left( \sum_{i=1}^k A(n, M_i) \mathbb{I}_{2\leq M_i \leq n}\right)
\end{align}

Because $A(n,m)$ scales linearly for large $n$, we can use the same bounding technique as in section \ref{app:inhom_bound} to show that $U(n) = O(n^{\beta})$ with $\beta < 2$ for large $n$. Therefore, we can conclude that $V(n) = O(n^{\beta})$ with $\beta < 2$ for large $n$, and so the growth rate $\frac{1}{n}N_m(T_n)$ converges in probability to the expected growth rate $a_m$ for large $n$.

\subsection{Typicality of Information Rate from Typicality of Growth Rates}
Using the convergence of the growth rates to their expected values, we can derive the convergence of the information rate to the entropy rate, using the following decomposition of the information content:
\begin{align}
\frac{1}{n} I(T_n) = \frac{S(n)}{n} + \sum_{m=2}^{n-1} \frac{N_m(T_n)}{n} S(m)
\end{align}

To establish that the information rate $\frac{1}{n} I(T_n)$ converges in probability to the expected entropy rate $h = \sum_{m=2}^{\infty} a_m S(m)$, we decompose the difference and analyze its limit. Note that the root splitting term $\frac{S(n)}{n} = \frac{1}{n} \log \binom{n+k-1}{k-1}$ decays as $O\left(\frac{\log n}{n}\right)$ and thus vanishes for large $n$. We are left to show that the sum of the random sub-tree contributions converges to the infinite series.

To do so, we fix an arbitrary integer $M \ge 2$. 
By the triangle inequality, we can bound the absolute difference by splitting it into three components - a finite head, a random tail, and a deterministic tail:
\begin{align}
\left\vert \sum_{m=2}^{n-1} \frac{N_m(T_n)}{n} S(m) - \sum_{m=2}^{\infty} a_m S(m) \right\vert \le \sum_{m=2}^{M} \left\vert \frac{N_m(T_n)}{n} - a_m \right\vert S(m) + \sum_{m=M+1}^{n-1} \frac{N_m(T_n)}{n} S(m) + \sum_{m=M+1}^{\infty} a_m S(m)
\end{align}

First, consider the deterministic tail (the third component). Because the series $\sum_{m=2}^{\infty} a_m S(m)$ converges to $h$, Eq.~\eqref{eq:h_rate1}, its infinite tail $\sum_{m=M+1}^{\infty} a_m S(m)$ can be made arbitrarily small by choosing a sufficiently large $M$.

Second, for the random tail (the second component), we examine its expected value:
\begin{align}
\mathbb{E}\left[ \sum_{m=M+1}^{n-1} \frac{N_m(T_n)}{n} S(m) \right] = \sum_{m=M+1}^{n-1} \frac{A(n,m)}{n} S(m)
\end{align}
Since we established in Appendix \ref{app:a_m}, that $\frac{A(n,m)}{n}$ is uniformly bounded by a dominating sequence $M_m=\frac{2}{m(m+1)}$ satisfying $\sum_{m=2}^{\infty} M_m S(m) < \infty$, the expectation of the random tail is bounded by $\sum_{m=M+1}^{\infty} M_m S(m)$, which can similarly be made arbitrarily small for large $M$. 
By Markov's inequality, since the expected value can be made arbitrarily close to zero for large $M$, the probability that the random tail exceeds any strictly positive value is also vanishingly small.

Finally, with $M$ fixed to a sufficiently large value to suppress the two tails, we turn to the finite head (the first component). Since we established in the variance analysis that $\frac{N_m(T_n)}{n}$ converges in probability to $a_m$ for each $m$ individually, and $M$ is finite, this finite sum converges in probability to $0$ as $n \to \infty$.

Because the tails can be made negligible by choosing a large $M$, and the finite head vanishes as $n \to \infty$, the entire difference converges in probability to $0$, which means that the information rate $\frac{1}{n} I(T_n)$ converges in probability to the expected entropy rate $h$ for large $n$.

\section{Subextensive Entropy} \label{app:subextensive}
\subsection{The logarithmic coefficient}
The sum $\sum_{m=2}^{n-1} p_n(m) H(m)$ can be approximated by the integral $\int _0^1 f(p) H(n p) dp$, where $f(p) = (k-1)(1-p)^{k-2}$ is the continuous limit of the splitting probability $p_n(m)$.
For $H(n) = A \log n$, we can show that this approximation - which includes replacing the sum by an integral, replacing $p_{n}(m)$ with its continuous limit $f(p)$, and extending integration limits - has an error of $O(\frac{\log n}{n})$ for large $n$.
Thus, $\sum_{m=2}^{n-1} p_n(m) \log n=\log n + O(1)$ for large $n$.

When we plug in $H(n)= A \log n$ into the recurrence relation, Eq.~\eqref{eq:H_recursive}, replace the splitting entropy $S(n)$ with its asymptotic leading term, $(k-1) \log n$, and use the above asymptotic approximation for the sum, we get: $A \log n = (k-1) \log n + k A \log n$, whose solution is $A=-1$.

\subsection{Hopf Bifurcation}
Apart from $s=-1$, roots that corresponds to a growing solution of $H(n)$, $\mathrm{Re}(s) <0$, must be complex - as explained in Appendix \ref{app:char_eq} - and so they correspond to solutions of the form $H(n) \sim n^{\mathrm{Re}(s)} \cos(\mathrm{Im}(s) \log n + \phi)$, which grow as a power law with log-periodic oscillations.

The characteristic equation $K(s)=1$ can be analyzed for continuous $k$ by formulating it in terms of the Gamma function:
\begin{align}
  \frac{\Gamma(k - s)}{\Gamma(1 - s)\Gamma(k + 1)} = 1
\end{align}
For low $k$ values, the only root that corresponds to a growing solution of $H(n)$, $\mathrm{Re}(s) <0$, is $s=-1$.
At a critical $k$ value, a pair of subdominant complex roots crosses the imaginary axis, $\mathrm{Re}(s)=0$, corresponding to oscillatory growing solutions -- a Hopf bifurcation.
Numerically, we find $k \approx 13.5$.


\addtocontents{toc}{\protect\setcounter{tocdepth}{3}}
\end{document}